# Optical detection of Mott and generalized Wigner crystal states in WSe$_2$/WS$_2$ moiré superlattices

Emma C. Regan[1,2,3]†, Danqing Wang[1,2,3]†, Chenhao Jin[1,4]†, M. Iqbal Bakti Utama[1,3,5], Beini Gao[1,6], Xin Wei[1,7], Sihan Zhao[1], Wenyu Zhao[1], Kentaro Yumigeta[8], Mark Blei[8], Johan Carlstroem[1,9], Kenji Watanabe[10], Takashi Taniguchi[10], Sefaattin Tongay[8], Michael Crommie[1,3,11], Alex Zettl[1,3,11], Feng Wang[1,3,11]*

[1] Department of Physics, University of California at Berkeley, Berkeley, California 94720, United States.

[2] Graduate Group in Applied Science and Technology, University of California at Berkeley, Berkeley, California 94720, United States.

[3] Material Science Division, Lawrence Berkeley National Laboratory, Berkeley, California 94720, United States.

[4] Kavli Institute at Cornell for Nanoscale Science, Ithaca, NY 14853, United States.

[5] Department of Materials Science and Engineering, University of California at Berkeley, Berkeley, California 94720, United States.

[6] Department of Physics, Huazhong University of Science and Technology, Wuhan 430074, China.

[7] School of Physics, University of Chinese Academy of Sciences, Beijing 100049, China.

[8] School for Engineering of Matter, Transport and Energy, Arizona State University, Tempe, Arizona 85287, United States.

[9] Department of Physics, Lund University, 221 00 Lund, Sweden.

[10] National Institute for Materials Science, 1-1 Namiki, Tsukuba, 305-0044, Japan.

[11] Kavli Energy NanoSciences Institute at University of California Berkeley and Lawrence Berkeley National Laboratory, Berkeley, California 94720, United States.

† These authors contributed equally to this work

* Correspondence to: fengwang76@berkeley.edu

**Abstract:**

Moiré superlattices are emerging as a new route for engineering strongly correlated electronic states in two-dimensional van der Waals heterostructures, as recently demonstrated in the correlated insulating and superconducting states in magic-angle twisted bilayer graphene and ABC trilayer graphene/boron nitride moiré superlattices[1–4]. Transition metal dichalcogenide (TMDC) moiré heterostructures provide another exciting model system to explore correlated quantum phenomena[5], with the addition of strong light-matter interactions and large spin-orbital coupling. Here we report the optical detection of strongly correlated phases in semiconducting $WSe_2/WS_2$ moiré superlattices. Our sensitive optical detection technique reveals a Mott insulator state at one hole per superlattice site ($\nu = 1$), and surprising insulating phases at fractional filling factors $\nu = 1/3$ and $2/3$, which we assign to generalized Wigner crystallization on an underlying lattice[6–9]. Furthermore, the unique spin-valley optical selection rules[10–12] of TMDC heterostructures allow us to optically create and investigate low-energy spin excited states in the Mott insulator. We reveal an especially slow spin relaxation lifetime of many microseconds in the Mott insulating state, orders-of-magnitude longer than that of charge excitations. Our studies highlight novel correlated physics that can emerge in moiré superlattices beyond graphene.

Moiré superlattices offer a general and powerful way to engineer correlated electronic states in van der Waals heterostructures. Consider a simplified but highly informative toy model: a two-dimensional (2D) electron gas in a periodic potential of periodicity *L*. The periodic potential leads to the formation of a set of minibands in the electron bandstructure. The on-site Coulomb potential *U* and the electronic bandwidth *W* of the lowest electronic miniband can be estimated as $U \sim \frac{e^2}{4\pi\varepsilon L}$ and $W \sim \frac{\hbar^2 k^2}{2m_e^*} \sim \frac{\hbar^2 \pi^2}{2m_e^* L^2}$, and the ratio U/W scales linearly with $m_e^*L$. Here $\varepsilon$ is the effective dielectric constant, and $m_e^*$ is the electron effective mass. Strong correlation (with U/W > 1) can be readily achieved with sufficiently large $m_e^*L$: for example, with a moiré superlattice (L ~ 10 nm) and an effective mass $m_e^* > 0.1m_0$ for $\varepsilon \sim 4\varepsilon_0$. If the periodic moiré potential is very strong, the electron bandwidth *W* will be additionally suppressed, further enhancing the correlation effects. The design criteria of large $m_e^*L$ can be satisfied in many moiré heterostructures. One successful instance is the ABC trilayer graphene/boron nitride moiré superlattice, which exhibits tunable Mott insulator, superconductor, and correlated Chern insulator states[3,4,13]. TMDC heterostructures represent another intriguing platform: the large effective mass ($m_e^* \sim 0.5m_0$) of TMDCs[14,15] can lead to particularly strong correlation effects, and the strong light-matter interactions[16,17] can enable optical detection and manipulation of the correlated quantum states of matter.

Here we report the observation of the Mott insulator state and generalized Wigner crystallization on an underlying lattice in semiconducting $WSe_2/WS_2$ moiré superlattices. Taking advantage of TMDCs' strong light-matter interactions, we optically detect both the quantum capacitance and electrical resistance of the moiré heterostructure while avoiding complications from very large contact resistances. A prominent example of strongly correlated electronic system is the Mott insulator at one hole per superlattice site (ν =1)[18,19], as illustrated in Fig. 1a. We show that the

Mott insulating state exists in WSe$_2$/WS$_2$ moiré superlattices up to 45 Kelvin and has an estimated Mott-Hubbard gap of ~10 meV, an order of magnitude larger than that in graphene moiré systems. Surprisingly, we also observe additional insulating states from generalized Wigner crystallization at fractional filling factors ν = 1/3 and 2/3. The emergence of these generalized Wigner crystal states necessitates an extended Hubbard model with not only on-site (short-range) but also inter-site (long-range) interactions[7,8]. In addition, the strong light-matter interaction and unique spin-valley selection rules allow us to optically create and detect different elementary excitations associated with the strongly correlated ground states in TMDC moiré heterostructures. We use circularly polarized light to generate a low-energy pure spin excitation, and we demonstrate an increased spin lifetime at the Mott insulating state.

In this work, we investigate correlated states in a TMDC heterostructure using a novel optically-detected resistance and capacitance (ODRC) technique. The large semiconductor bandgap in TMDCs leads to the formation of Schottky barriers at metal-TMDC junctions and correspondingly large contact resistance. This large contact resistance often hampers direct electrical transport measurements in TMDC heterostructures, particularly for low carrier doping and at low temperatures[20]. Our optical detection scheme avoids this difficulty associated with large contact resistance. For the ODRC measurements, we design a special device configuration with two regions (Fig. 1b): one half of the device has a local graphite top gate (region 1), and the other half does not (region 2). We vary the DC voltage on the local top gate (V$_{top}$) to continuously control the carrier doping in region 1, where the charge injection occurs with a time constant of ~ 1 s. We then add an AC excitation voltage ($\Delta \tilde{V}$) to the local top gate. For excitation frequencies higher than 10 Hz, the electrical contact is effectively frozen and the TMDC heterostructure is floated electrically (see Supplementary Information). In this case, the AC excitation voltage only leads to charge redistribution between region 1 and region 2 with no total charge change, and this charge redistribution dynamics depends on the quantum capacitance and resistance in the moiré system. We detect the resulting change of carrier concentration in region 2 ($\Delta \tilde{n}$) optically through the induced change in optical reflectivity $\Delta R/R$ at the intralayer exciton resonance (see Supplementary Information). The global graphite back gate is used to set the DC doping level of region 2 to optimize the exciton optical response to change in doping.

The AC electrical transport in the TMDC heterostructure can be modeled by an effective RC circuit shown in Fig. 1c. Here $C_1$ and $C_B$ are the geometric capacitance between the TMDC and the top and bottom gates in region 1, respectively, and $C_2$ is the TMDC-bottom gate capacitance in region 2. These geometric capacitance $C_i$ (i=1, 2, B) are set by $C_i = \frac{\varepsilon_0 \varepsilon_r A_i}{d_i}$, where $\varepsilon_r$ is the dielectric constant of the gate dielectric, and $A_i$ and $d_i$ denote the relevant capacitor area and separation. The parameters to be measured are $C_Q$ and $R$, which correspond to the doping-dependent quantum capacitance and resistance of the moiré superlattice in region 1, respectively. The induced optical contrast change ($\Delta R/R$) in region 2 upon an AC capacitive excitation in region 1 ($\Delta \tilde{V}$) can be obtained from the effective circuit model (see Supplementary Information) as

$$\frac{\Delta R}{R} = \alpha \Delta \tilde{n} = \frac{\alpha}{A_2 e} \Delta \tilde{V} \left( \frac{C_1}{C_1 + C_B} \right) \frac{1}{\frac{1}{C_{eff}} + i\omega R},$$

with

$$\frac{1}{C_{eff}} = \frac{1}{(C_1 + C_B)} + \frac{1}{C_2} + \frac{1}{C_Q}.$$

Here ω is the excitation frequency, $e$ is the electron charge, and $\alpha = (\frac{\Delta R}{R})/\Delta \tilde{n}$ is the optical detection responsivity in region 2, which is a constant for the fixed bottom gate voltage in our study. The frequency-dependent optical signal $\frac{\Delta R}{R}(\omega)$ allows us to extract the values of both $C_Q$ and R: At low excitation frequencies the resistance is negligible, so the optical signal probes the quantum capacitance $C_Q$, which is proportional to the density of states of the moiré heterostructure. At high modulation frequencies, both $C_Q$ and R contribute to the optical signal.

We focus our study on near-zero twist angle $WSe_2/WS_2$ heterostructure, which has a moiré superlattice with period of ~ 8 nm due to the ~ 4 % lattice mismatch between the $WS_2$ and $WSe_2$ monolayers. Figure 1d shows a schematic of the device: few-layer graphene is used for the gates and contact to the TMDC layers, and hBN is used at the top and bottom gate dielectrics ($\varepsilon_r = 4$, see Methods and Ref. [21] for fabrication details). Figure 1e shows the optical microscopy image of the final device, with contours highlighting the $WS_2$ and $WSe_2$ layers and the local graphite top gate. To verify the presence of the moiré superlattice, we examine the optical absorption spectrum of the heterostructure (Fig. 1f). It shows clear splitting of the $WSe_2$ A exciton, which is a signature of the moiré superlattice in the heterostructure[21].

Figure 2a shows the ODRC signals as a function of the hole doping of the $WSe_2/WS_2$ moiré superlattice in region 1. We use an AC excitation voltage with the peak-to-peak amplitude of 10 mV at 1 kHz and 30 kHz. When region 1 is near charge neutral ($V_{top} > 0.2$ V), the ΔR/R signal is small because no carriers are available to redistribute in the bandgap of $WSe_2$. When region 1 is hole doped ($V_{top} < 0.2$ V), charge redistribution occurs, leading to a large increase in signal. Interestingly, we observe a strong gap-like feature at -1 V (blue dashed line in Fig. 2a). From a capacitance model, we estimate the corresponding hole concentration to be 1.77 x $10^{12}$ cm$^{-2}$, which matches well with the density of one hole per moiré unit cell ($n_0 = 1.8 \times 10^{12}$ cm$^{-2}$). We also observe two sharp dips at -0.2 V and -0.6 V (orange and green dashed lines in Fig. 2a), which correspond to hole concentrations of $n = n_0/3$ and $n = 2n_0/3$, respectively. Additionally, a broad, weaker feature is observed at -2.25 V, corresponding to $n = 2n_0$. These features become stronger at higher excitation frequency of 30 kHz.

We extract numerical values for the doping-dependent $C_{eff}$ and R of the moiré heterostructure based on the effective AC circuit model and Eq. 1. We plot $C_{eff}$ and R as a function of carrier doping in Fig. 2b and 2c (grey lines), respectively. The optical responsivity of $\alpha = 1.4 \times 10^{-12}$ cm$^2$ is chosen so that $\frac{1}{C_{eff}} = \frac{1}{(C_1+C_B)} + \frac{1}{C_2}$ at high doping, where the quantum capacitance is much larger than the geometry capacitances and has negligible contribution. At $n = n_0$, $n = n_0/3$, and $n = 2n_0/3$, $C_{eff}$ decreases while the geometric capacitances remain unchanged (Fig. 2b). This decrease of $C_{eff}$ is due to a much smaller quantum capacitance $C_Q$, which results from significantly reduced density of states at these filling factors. At the same time, the electrical resistance shows marked increases at $n = n_0$, $n = n_0/3$, and $n = 2n_0/3$ (Fig. 2c). The simultaneous reduction of the density of state and large increase of the resistance indicate the emergence of insulating states at these fillings.

To quantitatively test our effective circuit model, we measure the frequency dependence of the ODRC signal at several representative hole-doping densities. Figure 2d displays the experimental data (symbols). We observe clear signal fall-off with increasing frequency, and data can be reproduced by the circuit model prediction (solid lines). The extracted effective capacitance and resistance at these fillings from the frequency dependence (black dots in Fig. 2c and 2d) agree well with the values extracted directly from the data in Fig. 2a.

Our results show that the WSe$_2$/WS$_2$ moiré heterostructure hosts insulating states with reduced density of states and increased resistance at n = n$_0$, n = n$_0$/3, and n = 2n$_0$/3. These features are completely absent in large twist angle WSe$_2$/WS$_2$ heterostructures (see Supplementary Information) and only emerge in the moiré superlattices. We assign the insulating state at n = n$_0$ to a Mott insulator (Fig. 2e)[18,19]. This corresponds to half filling of the moiré miniband because TMD heterostructure has a degeneracy of 2 from spin-valley locking[11]. Similar correlated insulating states have also been observed at n = n$_0$ in twisted bilayer graphene and ABC trilayer graphene/boron nitride moiré superlattices[1–4].

On the other hand, the observation of insulating states at n = n$_0$/3 and n = 2n$_0$/3 is quite surprising. Insulating states at fractional filling of the lattice sites have not been observed in other moiré superlattice systems and cannot be described as a Mott insulator or by a Hubbard model with only on-site repulsive interactions. We hypothesize that these insulating states at n = n$_0$/3 and n = 2n$_0$/3 correspond to generalized Wigner crystallization[6–9] of holes in the TMDC moiré superlattice. Figure 2e illustrates the real-space configurations of the generalized Wigner crystal states, where holes try to avoid not only double-occupation in one site, but also simultaneous occupation of adjacent sites. There are three degenerate Wigner crystallization configurations. The TMD moiré system spontaneously breaks the lattice translational symmetry due to the electron-electron interactions and condenses to one specific configuration with a $\sqrt{3} \times \sqrt{3}$ charge density wave pattern. The emergence of these generalized Wigner crystal states suggests that even the inter-site (long-range) interaction energy is larger than the moiré miniband bandwidth, confirming the very strong correlation in the TMDC moiré heterostructure.

We perform the ODRC measurements of the doping-dependent quantum capacitance and resistance of the TMDC moiré superlattices at different temperatures. Fig. 3a shows the extracted resistance for temperatures from 3 K to 70 K. The resistance peaks of the Mott insulator and generalized Wigner crystal states are observable up to temperatures of 45 K and 10 K, respectively. We estimate the Mott-Hubbard gap to be $\Delta \sim 10$ meV by fitting the resistance to a thermal activation function $\exp[-\Delta/(2k_BT)]$ for the Mott insulator state at n = $n_0$ (black dashed line in Fig. 3b). Due to the limited range exhibiting thermal activation behavior, the estimated Mott gap has relatively large uncertainty. It is difficult to estimate the size of the insulating gaps of the generalized Wigner crystal states from the experimental data, but they are likely to be 5-10 times smaller than the Mott insulator gap based on the temperature at which the generalized Wigner crystal signatures disappear.

The strong electron-correlation and light-matter-interaction in the heterostructure provides unique opportunities to optically investigate excited states from the correlated phases, such as low-energy charge and spin excitations. Charge excitations in Mott insulator systems have been intensively studied, featuring ultrafast decay dynamics (typically few picoseconds) from the holon-doublon recombination process[22–24]. On the other hand, the dynamics of pure spin-excitations have been difficult to explore. Here we directly measure the doping-dependent decay of a pure spin excitation by taking advantage of the unique spin-valley selection rules in the TMDC heterostructure[10–12]. We use the pump-probe scheme described in Ref.[25,26] to generate and probe the spin excitation in the moiré heterostructure. Specifically, a circularly polarized pump excitation is employed to selectively excite K-valley excitons composed of spin-up holes and electrons. The relaxation of the spin-polarized electrons and holes within about 100ns results in a residual spin polarization in the Mott insulator lower Hubbard band, as illustrated in Fig. 4a. We probe the evolution of the residual spin polarization through the pump-induced circular dichroism signal and the charge population through the pump-induced change in total absorption of a probe beam. Figure 4b shows the time evolution of the spin population at different hole densities. The doping-dependent spin lifetime, summarized as blue symbols in Fig. 4c, shows a prominent increase at the Mott insulator state ($n = n_0$) and reaches more than 8 us. In contrast, the lifetime of charge excitations (black symbols) is orders-of-magnitude shorter. The long-lived spin excitations from the Mott insulator state can provide important information about its spin configuration. It has been proposed that the Mott insulator state in the $WSe_2/WS_2$ moiré superlattice can host intriguing spin states such as quantum spin liquid[27,28]. However, further theoretical studies will be required to understand the experimentally observed spin dynamics in the Mott insulating state, which is beyond the scope of this paper.

Our results demonstrate the TMDC moiré heterostructures can host novel quantum correlated phases and offer an attractive platform for probing excited state and non-equilibrium dynamics of the correlated phases due to a unique combination of highly correlated electrons, strong light-matter interactions, and large spin-orbital effect in the system.

**Methods:**

Heterostructure preparation for optical measurements: We used a dry transfer method with a polyethylene terephthalate (PET) stamp to fabricate the $WSe_2/WS_2$ heterostructures[29]. Monolayer $WSe_2$, monolayer $WS_2$, few-layer graphene, and thin hBN flakes were first exfoliated onto Si substrates with a 90 nm $SiO_2$ layer. For aligned heterostructure, we used polarization-dependent second harmonic generation to determine the crystal axes of $WS_2$ and $WSe_2$[30,31]. We then used a PET stamp to pick up the few-layer graphene top gate, top hBN flake, the $WS_2$ monolayer, the $WSe_2$ monolayer, the few-layer graphene contact, the bottom hBN flake, and the few-layer graphene back gate in sequence. Between picking up $WS_2$ and $WSe_2$, we adjusted the angle of the PET stamp to ensure a near-zero twist angle between the flakes. The PET stamp with the above heterostructure was then stamped onto a clean Si substrate with 90 nm $SiO_2$. The PET and samples were heated to 60 °C during the pick up and to 130 °C for the stamp process. Finally, we dissolved the PET in dichloromethane overnight at room temperature. Afterwards, contacts (~75 nm gold with ~5 nm chromium adhesion layer) to the few-layer graphene flakes were made using electron-beam lithography and electron-beam evaporation.

ODRC measurements: A function generator (Rigol 1022Z) was used to generate the top gate voltage consisting of a DC offset, $V_{top}$, with small AC modulation, $\Delta \tilde{V}$. A voltage source (Keithley 2400) was used for the back gate voltage. A laser diode with center energy of 1.65 eV served as probe light. The diode energy was fine-tuned using a thermoelectric cooler, so that the probe energy was resonant with the lowest energy $WSe_2$ A exciton absorption peak in region 2. The reflected probe light was collected with an avalanche photo diode (Thorlabs APD 410A) and then analyzed using a lock-in amplifier that was locked to the function generator output.

Generation of optical pump-probe pulses with controlled time delay: Two electronic pulse generators (HP 8082A and HP 214B) were used to generate optical pump and probe pulses separately. Both pulse generators were triggered by the digital output of a data acquisition card, so the period and delay of the two triggering signals can be directly controlled with computer. The output electronic pulses with ~ 20 nanosecond pulse duration were then converted to optical pulses by two RF- coupled laser diode modules, with energies at 1.80 eV (pump) and 1.66 eV (probe), respectively. The pump and probe beams were focused at the sample with diameters of ~30 um and ~5 um, respectively. Their polarizations were set with linear polarizers and a shared quarter wave plate. The reflected probe light was collected by a photomultiplier tube. The pump-probe signal was analyzed using a lock-in amplifier at a ~ 2.5 kHz modulation frequency.

**References:**

1. Cao, Y. *et al.* Correlated insulator behaviour at half-filling in magic-angle graphene superlattices. *Nature* **556,** 80–84 (2018).
2. Cao, Y. *et al.* Unconventional superconductivity in magic-angle graphene superlattices. *Nature* **556,** 43–50 (2018).
3. Chen, G. *et al.* Evidence of a gate-tunable Mott insulator in a trilayer graphene moiré superlattice. *Nat. Phys.* **15,** 237–241 (2019).
4. Chen, G. *et al.* Signatures of tunable superconductivity in a trilayer graphene moiré superlattice. *Nature* (2019). doi:10.1038/s41586-019-1393-y
5. Wu, F., Lovorn, T., Tutuc, E. & MacDonald, A. H. Hubbard Model Physics in Transition Metal Dichalcogenide Moire Bands. *Phys. Rev. Lett.* **121,** 26402 (2018).
6. Wigner, E. On the Interaction of Electrons in Metals. *Phys. Rev.* **46,** 1002–1011 (1934).
7. Hubbard, J. Generalized Wigner lattices in one dimension and some applications to tetracyanoquinodimethane (TCNQ) salts. *Phys. Rev. B* **17,** 494–505 (1978).
8. Wu, C., Bergman, D., Balents, L. & Das Sarma, S. Flat Bands and Wigner Crystallization in the Honeycomb Optical Lattice. *Phys. Rev. Lett.* **99,** 70401 (2007).
9. Hiraki, K. & Kanoda, K. Wigner Crystal Type of Charge Ordering in an Organic Conductor with a Quarter-Filled Band: (D1-DCNQI)2Ag. *Phys. Rev. Lett.* **80,** 4737–4740 (1998).
10. Mak, K. F., He, K., Shan, J. & Heinz, T. F. Control of valley polarization in monolayer MoS2 by optical helicity. *Nat. Nanotechnol.* **7,** 494–498 (2012).
11. Xiao, D., Liu, G.-B., Feng, W., Xu, X. & Yao, W. Coupled Spin and Valley Physics in Monolayers of MoS2 and Other Group-VI Dichalcogenides. *Phys. Rev. Lett.* **108,** 196802 (2012).
12. Cao, T. *et al.* Valley-selective circular dichroism of monolayer molybdenum disulphide. *Nat. Commun.* **3,** 887 (2012).
13. Chen, G. *et al.* Tunable Correlated Chern Insulator and Ferromagnetism in Trilayer Graphene/Boron Nitride Moire Superlattice. *eprint arXiv:1905.06535* arXiv:1905.06535 (2019).
14. Kadantsev, E. S. & Hawrylak, P. Electronic structure of a single MoS2 monolayer. *Solid State Commun.* **152,** 909–913 (2012).
15. Fallahazad, B. *et al.* Shubnikov--de Haas Oscillations of High-Mobility Holes in Monolayer and Bilayer WSe2 Landau Level Degeneracy, Effective Mass, and Negative Compressibility. *Phys. Rev. Lett.* **116,** 86601 (2016).
16. Splendiani, A. *et al.* Emerging Photoluminescence in Monolayer MoS2. *Nano Lett.* **10,** 1271–1275 (2010).
17. Mak, K. F., Lee, C., Hone, J., Shan, J. & Heinz, T. F. Atomically thin MoS2: A new direct-gap semiconductor. *Phys. Rev. Lett.* **105,** 2–5 (2010).
18. Mott, N. F. The Basis of the Electron Theory of Metals, with Special Reference to the Transition Metals. *Proc. Phys. Soc. Sect. A* **62,** 416–422 (1949).
19. Imada, M., Fujimori, A. & Tokura, Y. Metal-insulator transitions. *Rev. Mod. Phys.* **70,** 1039–1263 (1998).
20. Allain, A., Kang, J., Banerjee, K. & Kis, A. Electrical contacts to two-dimensional semiconductors. *Nat. Mater.* **14,** 1195 (2015).
21. Jin, C. *et al.* Observation of moiré excitons in WSe2/WS2 heterostructure superlattices.

*Nature* **567,** 76–80 (2019).
22. Prelovs, P. & Lenarc, Z. Ultrafast Charge Recombination in a Photoexcited Mott-Hubbard Insulator. *Phys. Rev. Lett.* **111,** 16401 (2013).
23. Cuo, N. *et al.* Photoinduced transition from Mott insulator to metal in the undoped cuprates. *Phys. Rev. B* **83,** 125102 (2011).
24. Giannetti, C. *et al.* Ultrafast optical spectroscopy of strongly correlated materials and high-temperature superconductors: a non-equilibrium approach. *Adv. Phys.* **65,** 58–238 (2016).
25. Kim, J. *et al.* Observation of ultralong valley lifetime in WSe2/MoS2 heterostructures. *Sci. Adv.* **3,** e1700518 (2017).
26. Jin, C. *et al.* Imaging of pure spin-valley diffusion current in WS2-WSe2 heterostructures. *Science.* **360,** 893–896 (2018).
27. Law, K. T. & Lee, P. A. 1T-TaS as a quantum spin liquid. *Proc. Natl. Acad. Sci.* **114,** 6996–7000 (2017).
28. Grover, T., Trivedi, N., Senthil, T. & Lee, P. A. Weak Mott insulators on the triangular lattice: Possibility of a gapless nematic quantum spin liquid. *Phys. Rev. B* **81,** 245121 (2010).
29. Wang, L. *et al.* One-Dimensional Electrical Contact to a Two-Dimensional Material. *Science.* **342,** 614–617 (2013).
30. Kumar, N. *et al.* Second harmonic microscopy of monolayer MoS2. *Phys. Rev. B* **87,** 161403 (2013).
31. Li, Y. *et al.* Probing Symmetry Properties of Few-Layer MoS$_2$ and h-BN by Optical Second-Harmonic Generation. *Nano Lett.* **13,** 3329–3333 (2013).

**Acknowledgements**: We thank Shaowei Li for the helpful discussions, and we thank Conrad Stansbury and Caitlin Wang for their assistance with figure design. This work was supported primarily by the Director, Office of Science, Office of Basic Energy Sciences, Materials Sciences and Engineering Division of the US Department of Energy under contract number DE-AC02-05CH11231 (van der Waals heterostructures program, KCWF16). This work was also supported by the US Army Research Office under MURI award W911NF-17-1-0312. E.C.R. acknowledges support from the Department of Defense (DoD) through the National Defense Science & Engineering Graduate Fellowship (NDSEG) Program. C.J. acknowledges support from a Kavli Postdoctoral Fellowship. S.T. acknowledges support from NSF DMR-1552220 and 1838443.

**Author contributions:** E.C.R., D.W., and C.J. contributed equally. F.W. conceived the research. E.C.R, D.W., and C.J. carried out optical measurements. D.W., E.C.R, C.J., F.W. performed data analysis. E.C.R., D.W., B.G., X.W., M.I.B.U, S.Z., W.Z., J. C., M. C., A.Z. contributed to the fabrication of van der Waals heterostructures. K.Y., M.B., and S.T. grew $WSe_2$ and $WS_2$ crystals. K.W. and T.T. grew hexagonal boron nitride crystals. All authors discussed the results and wrote the manuscript.

**Competing interests:** The authors declare no competing interests.

**Data and materials availability:** The data that support the findings of this study are available from the corresponding author upon reasonable request.

**Corresponding author:** Correspondence and requests for materials should be addressed to fengwang76@berkeley.edu.

**Figures**

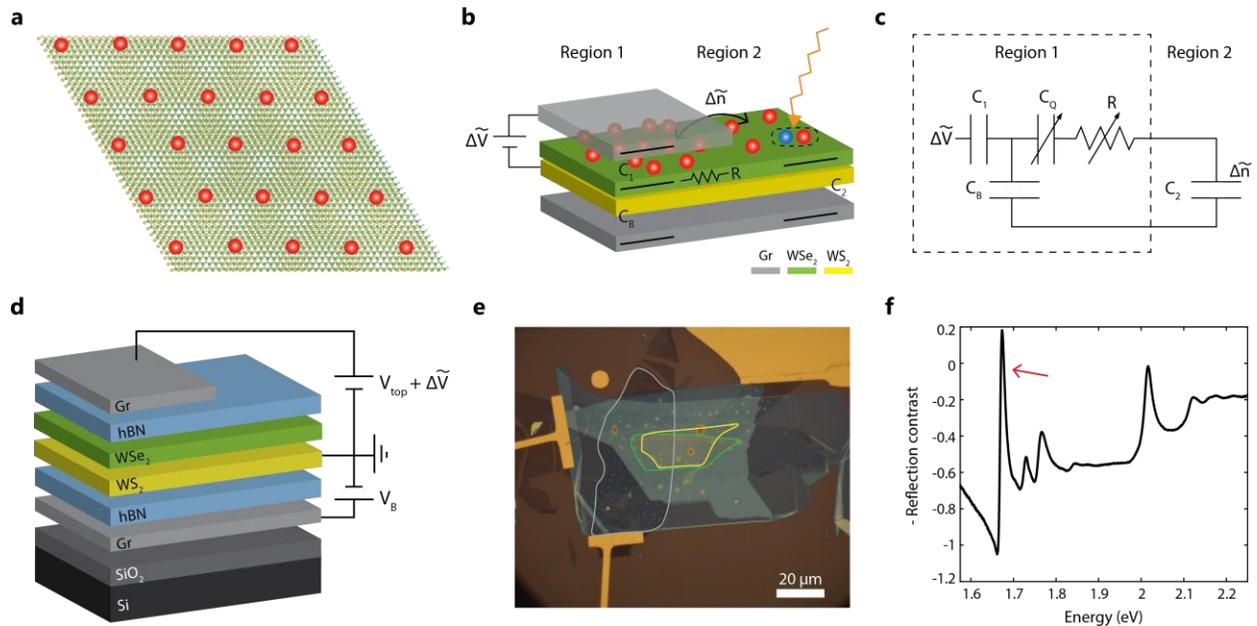

**Fig. 1 | Optically detected resistance and capacitance (ODRC) technique in WSe$_2$/WS$_2$ superlattice.** (**a**) Illustration of a Mott insulator state in a WSe$_2$/WS$_2$ moiré superlattice with one hole per superlattice unit cell. (**b** and **c**) Device schematic for an ODCR measurement in a WSe$_2$/WS$_2$ heterostructure (**b**), which includes a local top gate and global back gate. A small AC bias $\Delta\tilde{V}$ leads to charge redistribution between region 1 and region 2 ($\Delta\tilde{n}$), which is detected via the change in optical reflectivity of the WSe$_2$ exciton in region 2. This AC measurement can be modeled as an effective RC circuit (**c**), where the elements are shown schematically in (**b**). $C_1$, $C_B$, and $C_2$ are the illustrated geometric capacitances in the system, and R and $C_Q$ are the doping-dependent resistance and quantum capacitance of region 1 that we measure. (**d** and **e**) Side-view illustration (**d**) and optical microscope image (**e**) of a near-zero twist angle heterostructure. The graphite top gate, WS$_2$, and WSe$_2$ flakes are outlined in grey, yellow, and green, respectively. (**f**) Optical absorption spectrum of the heterostructure showing splitting of the WSe$_2$ A exciton into three prominent peaks, which is characteristic of intralayer moiré excitons in an aligned heterostructure. The ODRC measurements use a laser probe in resonance with the lowest energy exciton peak (red arrow).

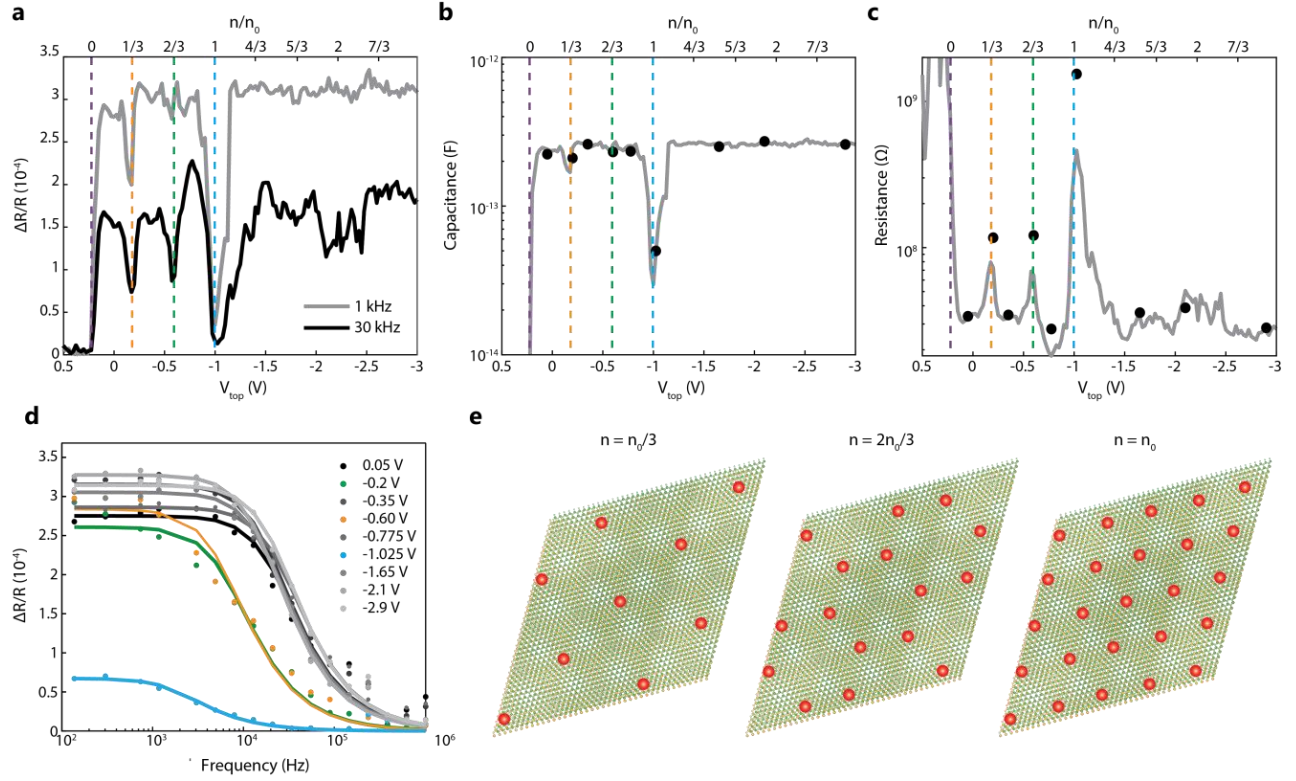

**Fig. 2 | Doping-dependent resistance and capacitance probed by ODRC.** (**a**) ODRC signal at 1 kHz (grey) and 30 kHz (black) from charge neutral to moderate hole doping. Strong gap-like features are observed at hole doping levels of $n = n_0/3$ (orange dashed line), $n = 2n_0/3$ (green dashed line), and $n = n_0$ (blue dashed line). The purple dashed line corresponds to $n = 0$. (**b-d**) Extracted capacitance $C_{eff}$ (**b**) and resistance (**c**) of region 1 from the data in (**a**) (grey curves) and from the frequency-dependent ODRC signal (**d**) at representative doping levels (black dots). The decreased capacitance and increased resistance indicate emerging insulating states at $n = n_0/3$, $n = 2n_0/3$, and $n = n_0$. All measurements are done at 3 K. (**f**) Illustrations of generalized Wigner crystal ($n = n_0/3$, $n = 2n_0/3$) and Mott insulator states ($n = n_0$) in a WSe$_2$/WS$_2$ moiré superlattice.

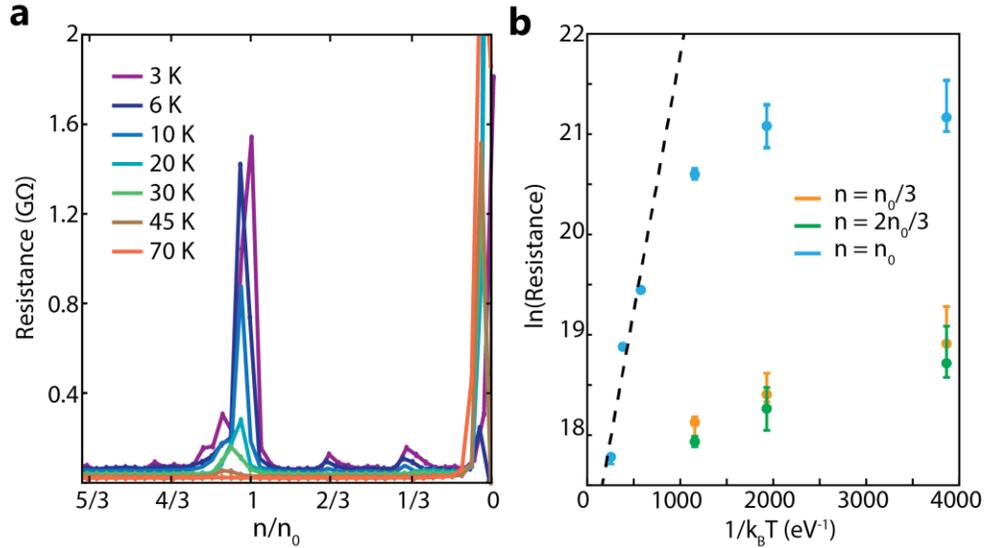

**Fig. 3 | Temperature dependence of Mott and generalized Wigner crystal states.** (**a**) Extracted resistance from ODRC measurements taken at a range of temperatures between 3 K and 70 K. The Mott insulator state at $n = n_0$ is observable up to 45 K, and the generalized Wigner crystal states at $n = n_0/3$ and $n = 2n_0/3$ persist until 10 K. (**b**) Plot of ln(Resistance) versus $1/k_B T$ for Mott (blue) and generalized Wigner crystal states (orange and green). Error bars correspond to the estimated uncertainty in the extracted resistance values. The estimated thermal activation gap for the Mott state is ~ 10 meV, which is found by fitting the data in (**b**) to a thermal activation function (black dashed line).

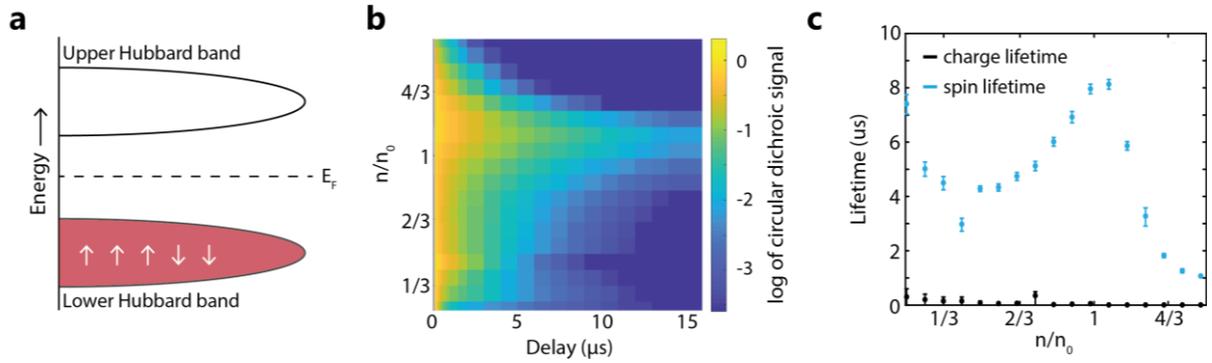

**Fig. 4 | Optical investigation of low-energy spin excitation dynamics of a $WSe_2/WS_2$ Mott insulator.** (**a**) A circularly polarized pump pulse selectively excites K-valley excitons with spin-up electrons and holes in the moiré heterostructure. After a fast charge recombination process, an excess of spin-polarized holes remains in the lower Hubbard band. The evolution of this low-energy spin excitation can be measured by a second probe pulse through pump-induced circular dichroism signals. (**b**) Doping dependent decay dynamics of the optically generated pure spin excitations. (**c**) Summary of the spin (blue) and charge (black) lifetimes as a function of hole doping. The spin relaxation slows down markedly near $n = n_0$, with a lifetime as long as 8 μs. On the other hand, the total charge population decays quickly for all doping levels.

**Supplementary Information for**

**Optical detection of Mott and generalized Wigner crystal states in WSe$_2$/WS$_2$ moiré superlattices**

Emma C. Regan†, Danqing Wang†, Chenhao Jin†, M. Iqbal Bakti Utama, Beini Gao, Xin Wei, Sihan Zhao, Wenyu Zhao, Kentaro Yumigeta, Mark Blei, Johan Carlstroem, Kenji Watanabe, Takashi Taniguchi, Sefaattin Tongay, Michael Crommie, Alex Zettl, Feng Wang[*]

† These authors contributed equally to this work

* Correspondence to: fengwang76@berkeley.edu

**S1. WSe$_2$ A exciton absorption in region 2**

**S2. Low-frequency behavior of ODRC signal**

**S3. Effective AC circuit model**

**S4. ODRC signal in large twist angle WSe$_2$/WS$_2$ heterostructure**

## S1. WSe$_2$ A exciton absorption in region 2

The lowest-energy WSe$_2$ A exciton in region 2 is used to measure the electrical properties of region 1 as the local top gate is tuned. To be a reliable probe, the exciton in region 2 should only respond to charge redistribution due to the modulation voltage $\Delta\tilde{V}$, but not to the DC bias applied to the local top gate, $V_{top}$. We measure the optical spectrum of the lowest-energy WSe$_2$ A exciton (between 1.6 to 1.74 eV) while sweeping $V_{top}$ from 0.5 V to -3.5 V (Fig. S1a). We record the spectrum 15 s after changing $V_{top}$ to ensure the contact injects charge. We observe almost no change in the spectrum. Therefore, the static local top gate does not influence the hole concentration in region 2, and it remains a stable probe for all $V_{top}$ used.

We measure the absorption spectrum as a function of the carrier density in region 2 by varying the global back gate voltage. Figure S1b shows the absorption spectra when the back gate voltage is tuned in a small range around -1 V. For the ODCR measurement, the modulation voltage $\Delta\tilde{V}$ is typically set to 10-25 mV, so the redistributed charges correspond to a small back gate voltage change of <25 mV. Within this range, the exciton resonance (1.673 eV) shows monotonic and linear change with carrier concentration. Therefore, α is a constant for our choice of back gate voltage.

We estimate an optical detection responsivity α of ~ 1.4 x 10$^{-12}$ cm$^2$. The noise of the ODCR signal is at ~ 2 x 10$^{-6}$ level in our lock-in measurement for 3 s average time. It allows us to detect a carrier density change in region 2 as small as 10$^6$ cm$^{-2}$ with optical detection.

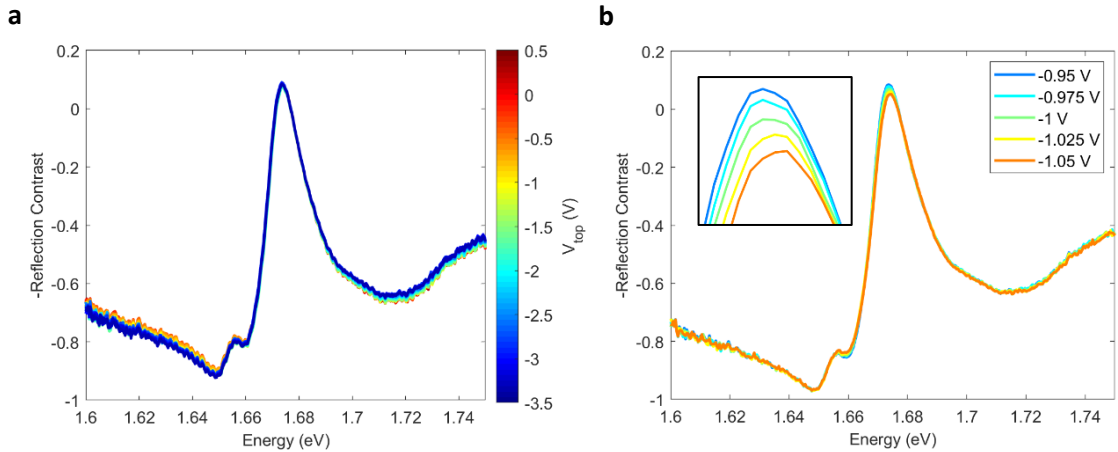

Fig S1. **a**. Reflection contrast spectra for the lowest-energy WSe$_2$ A exciton resonance in region 2 when the local top gate voltage $V_{top}$ is tuned from 0.5 V to -3.5 V. Region 2 is not affected when the hole concentration is tuned in region 1 by $V_{top}$. **b.** Reflection contrast spectra for the WSe$_2$ A exciton in region 2 when the global back gate is tuned from -0.95 V to -1.05 V. Inset shows zoomed-in view of the exciton peak. The spectral change is monotonic and approximately linear with carrier concentration.

## S2. Low-frequency behavior of ODRC signal

Figure S2 shows the frequency-dependent ODCR signal at $V_{top}$ = -1.6 V (i.e. away from any features) for modulation frequencies between 0.05 Hz to 137 Hz. The carrier injection through the graphite contact has a characteristic time constant of ~ 1s. At the lowest modulation frequency (0.05 Hz), the graphite contact can efficiently inject charge in response to $\Delta \tilde{V}$. As a result, the carrier density in region 2 remains a constant and the overall ODCR signal is negligible. At 1 Hz, the ODCR signal is partially reduced compared with higher frequency responses because the contact can inject some charge in response to $\Delta \tilde{V}$. For frequencies higher than ~10 Hz, the contact becomes frozen. As a result, the heterostructure is effectively floated and the ODCR signal reaches its typical low-frequency value.

We also note that the ODCR signal is linear with $\Delta \tilde{V}$.

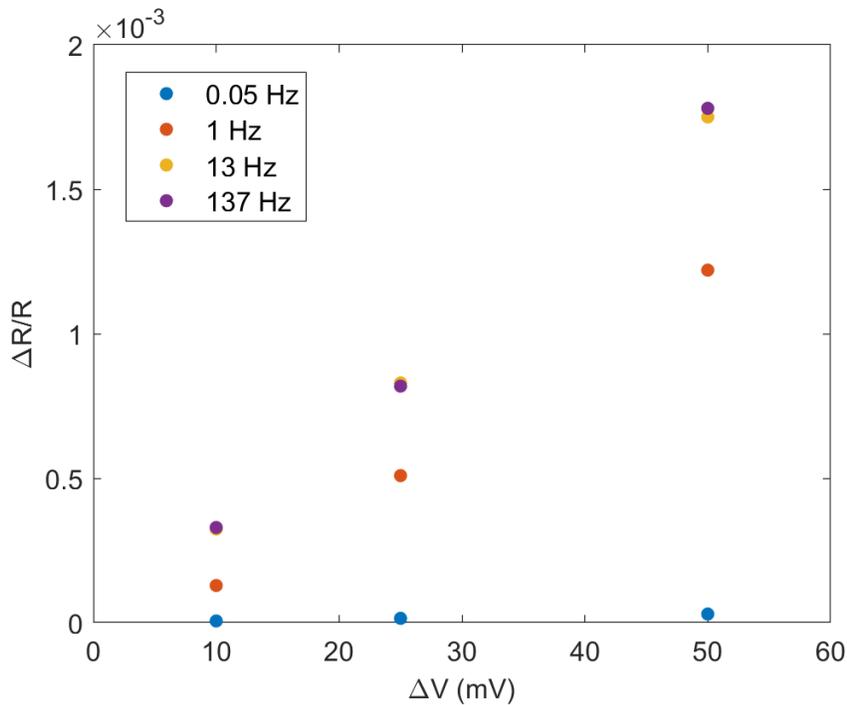

Figure S2. ODCR signal measured at very low frequency for a range of modulation voltages.

## S3. Effective AC circuit model

We use an effective AC RC circuit, shown in Figure 1c, to model the AC electrical transport in the TMDC heterostructure. $C_1$ and $C_B$ are the geometric capacitances between the TMDC and the top and bottom gates in region 1, respectively, and $C_2$ is the geometric capacitance between the TMDC and the bottom gate in region 2. The geometric capacitances are defined by $C_i = \frac{\varepsilon_0 \varepsilon_r A_i}{d_i}$, where $\varepsilon_r$ is the dielectric constant of the gate dielectric, and $A_i$ and $d_i$ denote the relevant capacitor area and separation. $C_Q$ and $R$ are the quantum capacitance and resistance of region 1, respectively. The ODCR signal measures the modulation induced optical contrast change in region 2, which can be understood as the optical response to an effective voltage change on region 2 ($\Delta V_2$) due to the AC capacitive excitation in region 1 ($\Delta \tilde{V}$):

$$\Delta V_2 = \Delta \tilde{V} \left( 1 - \frac{\frac{1}{i\omega C_1}}{\frac{1}{i\omega C_1} + \frac{1}{i\omega C_B + \frac{1}{\frac{1}{i\omega C_Q} + \frac{1}{i\omega C_2} + R}}} \right) \frac{\frac{1}{i\omega C_2}}{\frac{1}{i\omega C_Q} + R + \frac{1}{i\omega C_2}},$$

which can be simplified to:

$$\Delta V_2 = \Delta \tilde{V} \left( \frac{C_1}{C_1 + C_B} \right) \frac{\frac{1}{C_2}}{\frac{1}{(C_1 + C_B)} + \frac{1}{C_Q} + \frac{1}{C_2} + i\omega R}.$$

For simplicity, we define an effective capacitance as

$$\frac{1}{C_{eff}} = \frac{1}{(C_1 + C_B)} + \frac{1}{C_2} + \frac{1}{C_Q},$$

so

$$\Delta V_2 = \Delta \tilde{V} \left( \frac{C_1}{C_1 + C_B} \right) \frac{\frac{1}{C_2}}{\frac{1}{C_{eff}} + i\omega R}.$$

The effective voltage change leads to a total charge change $\Delta \tilde{Q}$ in region 2:

$$\Delta \tilde{Q} = \Delta V_2 C_2 = \Delta \tilde{V} \left( \frac{C_1}{C_1 + C_B} \right) \frac{1}{\frac{1}{C_{eff}} + i\omega R},$$

and corresponding change in hole density:

$$\Delta \tilde{n} = \frac{\Delta \tilde{Q}}{A_2 e} = \frac{1}{A_2 e} \Delta \tilde{V} \left( \frac{C_1}{C_1 + C_B} \right) \frac{1}{\frac{1}{C_{eff}} + i\omega R}$$

where e is the electron charge. The ODCR signal, $\Delta R/R$, is related to $\Delta \tilde{n}$ by the optical detection responsivity $\alpha$, which describes the change in the optical absorption for a given change in doping:

$$\frac{\Delta R}{R} = \alpha \Delta \tilde{n} = \frac{\alpha}{A_2 e} \Delta \tilde{V} \left( \frac{C_1}{C_1 + C_B} \right) \frac{1}{\frac{1}{C_{eff}} + i\omega R}.$$

The responsivity $\alpha$ is set to 1.4 x 10$^{-12}$ cm$^2$, so that $\frac{1}{C_{eff}} = \frac{1}{(C_1 + C_B)} + \frac{1}{C_2}$ at high doping, where the quantum capacitance is much larger than the geometry capacitances and has negligible contribution. This value of $\alpha$ also is also consistent with the value determined by the doping-dependent optical absorption of region 2.

## S4. ODRC signal in large twist angle WSe$_2$/WS$_2$ heterostructure

We measure the ODCR signal for large twist angle WSe$_2$/WS$_2$ heterostructure In this device, the monolayer WSe$_2$ and WS$_2$ flakes are intentionally misaligned, and the absorption spectrum is characteristic of a large twist angle heterostructure. The signal from misaligned heterostructure also shows sharp increase when doped below the bandgap (red curve in Fig. S3a), indicating that charge redistribution occurs. However, the signal is largely flat and does not show any clear dips corresponding to insulating states, in sharp contrast with the aligned case (blue curve). This observation is consistent with our conclusion that the insulating states in an the aligned heterostructure are Mott and generalized Wigner states in the moiré superlattice, which is not present in a large twist angle heterostructure. Figure S3b shows the ODCR signal at several representative frequencies, showing a characteristic RC circuit fall-off with increasing frequency. No additional feature is observed in the hole-doping region up to the frequency of 1 MHz, further confirming the absence of insulating states. The overall lower resistance in the misaligned device may be due to the difference in back gate doping used in the two measurements.

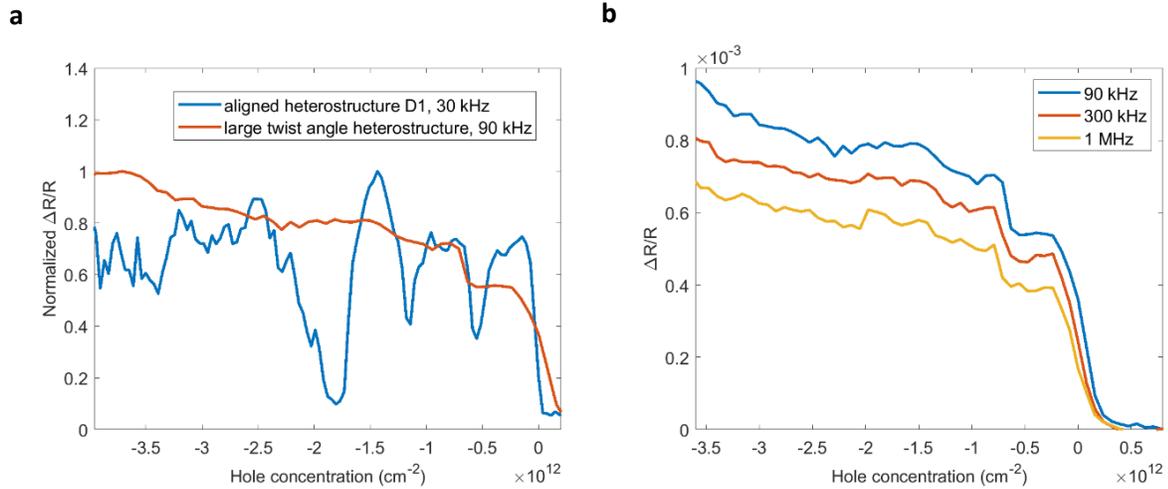

Fig S3. **a**. Normalized ΔR/R for a large twist angle heterostructure (red) and aligned heterostructure (blue). The misaligned heterostructure does not show any insulating features. **b.** The frequency dependence of the large twist angle signal, showing a characteristic RC circuit fall-off with increasing frequency.